\documentclass[10pt,twoside,notitlepage,twocolumn]{article}
\usepackage{letter}

\newcommand\email[1]{%
	\begingroup
	\renewcommand\thefootnote{}\footnote{#1}%
	\addtocounter{footnote}{-1}%
	\endgroup
}
\begin{document}
\allsectionsfont{\sffamily}

%
% *** FRONT MATTER
%

% *** title *** %
\title{%
Molecular interfacial rheology:
Lipid membrane shear viscosity%
}

% *** short title *** %
\shorttitle{Molecular interfacial rheology: Lipid membrane shear viscosity}

% *** authors *** %
\author{Zhi-Xun Xu$\,^{\dag,}$}
\author{Amaresh Sahu$\,^{\ddag,}$}

% *** short authors *** $
\shortauthor{Z.-X.\ Xu and A.\ Sahu}

% *** affiliations *** %
\affil{McKetta Department of Chemical Engineering, University of Texas, Austin TX 78712, USA}

% *** date *** %
\date{4 September 2026}

%
% *** COMPILE FRONT MATTER
%

\twocolumn[
	\begin{@twocolumnfalse}
		% *** Make Title *** %
		\maketitle
		% *** Abstract *** %
		\begin{abstract}
			%\noindent\textsf{\textbf{Abstract.}}
			%
%
%
We develop a method to extract the shear viscosity of a lipid membrane from equilibrium molecular dynamics simulations.
The method characterizes the rheology of general interfacial systems embedded in three-dimensional media; we term it \textit{molecular interfacial rheology.}
In our simulations the planar bilayer and surrounding water are confined between solid, parallel walls.
Following Onsager's regression hypothesis, membrane and water fluctuations are assumed to relax according to the coupled continuum-mechanical equations governing the confined system---which predict that the membrane transverse velocity autocorrelation function (\tvacf) decays exponentially, at a rate set by the membrane and water viscosities.
The measured \tvacf, however, exhibits damped oscillations followed by a slowly decaying tail.
We reconcile these behaviors using the Mori--Zwanzig formalism, and extract the wavevector-dependent membrane viscosity from the time-integral of the \tvacf.
Results from theory and simulations agree over a decade of wavevectors, and extrapolating to long wavelengths yields shear viscosities ranging from 0.064 to 0.18 pN$\mk \cdot \mk$\textmu s$/$nm across two representative single-component, fluid-phase bilayers.
Our results are corroborated by nonequilibrium simulations where a spatially varying in-plane body force is applied to lipid molecules, thus validating the framework of molecular interfacial rheology.
		\end{abstract}
		\vskip 1.9em
	\end{@twocolumnfalse}
]
\thispagestyle{empty}

\email{$^\dag \,$\href{mailto:zhixunxu@utexas.edu}{\texttt{zhixunxu@utexas.edu}}}
\email{$^\ddag \,$\href{mailto:asahu@che.utexas.edu}{\texttt{asahu@che.utexas.edu}}}

% *** CONTENT *** %
\normalsize
\vspace{-15pt}

%
% *** Introduction
%
\smallskip\noindent
\textsf{\textbf{Introduction.}}---%
In multiphase systems, the interfaces between phases contribute to, and in some cases dictate, the stability and dynamics of the overall system \cite{stone-jfm-2010, edwards-brenner, fuller-arcbe-2012}.
There are accordingly many investigations into the rheology of interfaces, e.g.\ in the study of liquid or solid foams \cite{stone-jpcm-2002, durand-stone-prl-2006, prudhomme}, emulsions \cite{smith-1976, russel-1989, utada-s-2005}, thin films \cite{mysels, bhamla-po-2017}, and soluble or insoluble surfactants \cite{rosen-surfactants, manikantan-jfm-2020}.
Interfaces are also essential in biology.
At cellular and sub-cellular scales, the most relevant biological interface is the lipid bilayer: a two-molecule-thick material surface exhibiting
out-of-plane elasticity and in-plane fluidity.
While the in-plane, intramembrane shear viscosity $ \zeta $ governs the evolution of lipid flows and membrane shape changes \cite{sahu-pre-2020}, it remains difficult to measure experimentally---with reported values spanning several orders of magnitude \cite{honerkamp-prl-2013, nagao-prl-2021, faizi-bpj-2022, suja-prf-2025}.
Calculating $ \zeta $ from molecular dynamics (\md) simulations is also challenging.
Difficulties arise from
($i$) discrepancies \cite{camley-jcp-2015, venable-jpcb-2017} between hydrodynamic calculations \cite{saffman-pnas-1975, saffman-jfm-1976, hughes-jfm-1981, stone-jfm-1998, petrov-bpj-2008, petrov-bpj-2012} and the periodic boundary conditions (\pbcs) used in simulations \cite{den-otter-bpj-2007, vogele-prl-2018, zgorski-jctc-2019},
($i \mkn i$) ad-hoc adaptations of established three-dimensional results \cite{zgorski-jctc-2019, fitzgerald-bpj-2023}, and
($i \mkn i \mkn i$) the requisite large shear rates in nonequilibrium simulations \cite{den-otter-bpj-2007, zgorski-jctc-2019}.
New approaches are thus needed to calculate the membrane shear viscosity from \md simulations.

%% begin figure %%
\begin{figure}[!b]
    \centering
    \vspace{-12pt}
    \includegraphics[width=0.92\columnwidth]{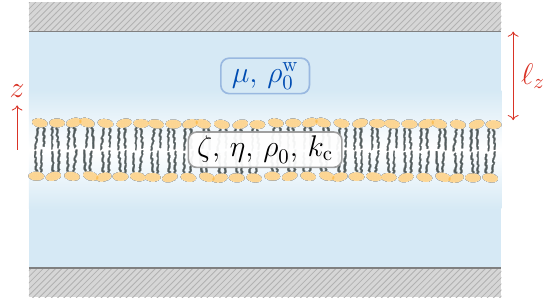}
    \vspace{-3pt}
    \caption{%
        Schematic of the membrane--water system, surrounded by solid walls, in \md simulations.
        Membrane and water parameters are described in the text.
        The bilayer is centered on the $ x $--$ y $ plane, and $ \ellz $ is the water thickness above and below the membrane.
        Periodic boundary conditions are used in the in-plane directions.%
    }
    \label{fig_config}
\end{figure}
%% end figure %%

\begin{figure}[!b]
    \centering
    \vspace{-14pt}
    \includegraphics[width=0.92\linewidth]{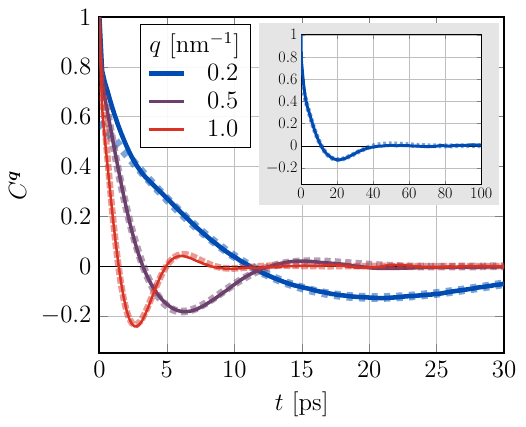}
    \vspace{-8pt}
    \caption{%
        Normalized \tvacfs from \md simulations of a \dopc bilayer in the fluid phase, at 25$^\circ$ C.
        The lateral size of the simulation cell is
        $ \ellx = \elly \approx 62.5 $ nm,
        for which
        $ q \ge 2 \pi / \ellx \approx $ 0.1 nm$^{-1}$.
        The \md data (thin solid lines) are observed to be well-approximated by
        $
            [
                A^q \cos (\gamma^q t)
                + B^q \sin (\gamma^q t)
            ] \exp \{- \alpha^{\mkn q} t \}
        $,
        as shown by the thick, transparent dashed lines.
        %The scaling of $ \gammaq $ with $ q $, shown in Fig.\ \ref{fig_gamma_q}, suggests why oscillations arise and $ \Cq $ contradicts the hydrodynamic prediction of exponential decay \eqref{eq_Ch}.
        (inset) Data for
        $ q \approx 0.2 $ nm$^{-1}$
        are shown over a longer time window.
    }
    \label{fig_C_t}
\end{figure}

In this Letter, we develop a method to calculate interfacial shear viscosities via equilibrium \md simulations---hereafter referred to as \textit{molecular interfacial rheology} (\mir).
Our techniques are general, and can be applied to any molecular interface.
Here, we focus on lipid bilayers as a complex model system due to the aforementioned challenges in measuring the shear viscosity.
We avoid hydrodynamic correlations between membrane sheets, which arise from the use of \pbcs in the normal direction \cite{camley-jcp-2015, venable-jpcb-2017}, by simulating the bilayer and surrounding water between parallel walls (see Fig.\ \ref{fig_config}).
Following Onsager's regression hypothesis \cite{onsager-pr-1931-i, onsager-pr-1931-ii}, membrane and water fluctuations are assumed to evolve according to the hydrodynamics of the confined system.
We thus expect a mode with initial velocity
$ \bmv \sim v^{}_0 \mk \cos (q \mk y) \mk \bmex $
to decay exponentially, at a rate that depends on $ \zeta $.
In simulations, however, such modes exhibit damped oscillations, as shown in Fig.\ \ref{fig_C_t}.
The differences in hydrodynamic and microscopic responses are bridged through the Mori--Zwanzig formalism and language of generalized hydrodynamics \cite{mori-pr-1958, zwanzig-jcp-1960, zwanzig-pr-1961, mori-ptp-1965, evans-morriss, hansen}.
In what follows, we present our analysis of dioleoylphosphatidylcholine (\dopc) bilayers at 25$^\circ \,$C, for which
$ \zeta^{}_{\dopc} = 0.184 \pm 0.006 $ pN$\mk \cdot \mk$\textmu s$/$nm.
The supplemental material (\sm) \cite{supplemental} contains the corresponding analysis of dipalmitoylphosphatidylcholine (\dppc) membranes at 67$^\circ \,$C, with
$ \zeta^{}_{\dppc} = 0.064 \pm 0.005 $ pN$\cdot$\textmu s$/$nm;
both bilayers are in the fluid phase.

% begin figure
% end figure

%
% *** Continuum response motivates MD predictions
%
\smallskip\noindent
\textsf{\textbf{Continuum response motivates MD predictions.}}---%
We begin with the linearized hydrodynamics of a membrane surrounded by water, under the assumption that velocity and density fluctuations are small.
Our derivations are detailed in Ref.\ \cite{xu-fluids}, and only summarized here.
%
%The water above ($ + $) and below ($ - $) the membrane is incompressible, with velocity components $ \vpm_j $ satisfying
%$ \vpm_{j, \mk j} = 0 $.
%Here and from now on, $ j $ and other Roman indices span $ \{ 1, 2, 3 \} $, Greek indices span $ \{ 1, 2 \}, $ commas denote partial differentiation, and repeated indices are summed over.
%Water is governed by the time-dependent Stokes equation
%$
%    \rhowz \mk \vpm_{j, \mk t}
%    = \mu \mk \vpm_{j, \mk k k}
%    - \ppm_{, j}
%$,
%where $ \ppm $ is the pressure field, $ \rhowz $ is the mass density, and $ \mu $ is the water shear viscosity.
%
We model the lipid bilayer as a two-dimensional (2D) surface of zero thickness, following prior theoretical developments \cite{hu-pre-2007, arroyo-pre-2009, kranthi-bmmb-2012, sahu-pre-2017}.
The continuity equation of a compressible membrane is given by
$
    \partial_t \mk \rho
    \mk + \rhoz \mkn \bmnablapara \bmcdot \bmv
    = 0
$,
where $ \bmv $ is the in-plane velocity and $ \bmnablapara $ is the 2D gradient operator in the $ x $--$ y $ plane.
Here $ \rhoz + \rho $ is the total mass density per unit area: $ \rhoz $ is the average areal density and $ \rho $ is the density perturbation, with
$ \lvert \mk \rho \mk \rvert \ll \rhoz $.
The local form of the in-plane momentum balance is given by
$
    \rhoz \mk \partial_t \mk \bmv
    = \zeta \mk \nablapara^2 \bmv
    + \eta \mk \bmnablapara (\bmnablapara \bmcdot \bmv)
    - \cst \mk \bmnablapara \rho
    + \bmf
$,
where $ \zeta $ and $ \eta $ are respectively the membrane shear and dilational viscosities,
$ \cs := (\kc / \rhoz)^{1/2} $
is the speed of sound with $ \kc $ the areal compressibility modulus, and $ \bmf $ is the in-plane force per area on the membrane by the surrounding fluid.
Fluid layers above and below the bilayer have the same thickness $ \ellz $ (see Fig.\ \ref{fig_config}), for which the membrane height $ h $ is decoupled from $ \bmv $ and $ \rho $---and so its evolution is not discussed here \cite{xu-fluids}.
The water surrounding the membrane is described by the incompressible Stokes equations.
We assume no-slip and traction-free boundary conditions, respectively, at water--membrane and water--wall interfaces.
The latter reflects our use of structureless walls in \md simulations, which by construction exert no tangential forces on the fluid \cite[\S5.1]{supplemental}.

We solve for the coupled membrane and fluid behavior by decomposing unknowns into normal modes.
The in-plane membrane velocity $ \bmv(\bmx, t) $, where $ \bmx $ is the in-plane position, is given by
$
    \bmv (\bmx, t)
    = \sum_{\bmq} \hatbmvq \mkn (t) \, \expq 
$.
Here $ \bmq $ is the in-plane wavevector, with magnitude $ q $.
Each velocity component $ \hatbmvq $ is projected in the longitudinal and transverse directions, which are respectively parallel and perpendicular to $ \bmq $.
To this end, we define
$ \bmuqpar := (q^{}_x, q^{}_y) / q $
and
$ \bmuqperp := (-q^{}_y, q^{}_x) / q $,
for which
$
    \hatbmvq
    \mk = \mk \hatvqpar  \, \bmuqpar
    \mk + \mk \hatvqperp \, \bmuqperp
$ \cite{parallel}.
The transverse modes $ \hatvqperp $ are decoupled from the longitudinal modes $ \hatvqpar $ and density modes $ \hatrhoq $, and evolve in time according to
\begin{equation} \label{eq_vqperp}
    \dd{\hatvqperp (t)}{t}
    \, + \, \mk \xiq \, \hatvqperp (t)
    \ = \ 0
    ~.
\end{equation}
The linearized hydrodynamic equations accordingly predict transverse modes decay exponentially at the rate $ \xiq $, which captures the drag from the membrane and water and is given by
\begin{equation} \label{eq_xiq}
    \xiq
    \mk := \, \dfrac{1}{\rhoz} \, \Big(
        \zeta \mk q^2
        \, + \, 2 \mk \mu \mk q \mk \tanh \big( \ellz \mk q \big)
    \Big)
    ~.
\end{equation}
Note that the hyperbolic tangent in Eq.\ \eqref{eq_xiq} results from traction-free wall--water interfaces; other functions enter for no-slip or finite-slip boundaries \cite{xu-fluids}.
%From Eq.\ \eqref{eq_xiq}, we recognize a fundamental difficulty of analyzing interfacial transport coefficients with methods that take the long-wavelength limit:
%when
%$ q \rightarrow 0 \mk $,
%$
%    \xiq
%    \sim (2 \mk \mu \mk q / \rhoz) \coth (\ellz q)
%    \sim 2 \mk \mu / (\rhoz \mk \ellz)
%$
%and the membrane viscosity does not contribute to the drag.

In \md simulations, spontaneous velocity fluctuations arise from thermal motions.
While the dynamics of each fluctuation appears stochastic, we follow Onsager and assume the average relaxation of fluctuations is governed by the continuum equations \cite{onsager-pr-1931-i, onsager-pr-1931-ii}.
The normalized transverse velocity autocorrelation function (\tvacf), defined as
$
    \Cq (t)
    := \langle \mk \hatvqperp (t) \, \hatvnqperp (0) \mk \rangle
    \, / \, \langle \mk \hatvqperp (0) \, \hatvnqperp (0) \mk \rangle
$,
is thus expected to satisfy [cf.\ Eq.\ \eqref{eq_vqperp}]
\begin{equation} \label{eq_Ch}
    \dd{\Cqh (t)}{t}
    \, + \, \xiq \mk \Cqh (t)
    \, = \, 0
    ~.
\end{equation}
Here $ \langle \, \bmcdot \, \rangle $ denotes an ensemble average and the subscript `h' indicates that Eq.\ \eqref{eq_Ch} is a hydrodynamic prediction.
The membrane shear viscosity could then be calculated by measuring $ \Cq (t) $ in \md simulations, extracting $ \xiq $ as the rate of exponential decay, and determining $ \zeta $ by inverting Eq.\ \eqref{eq_xiq}---in which all other parameters are known.

\begin{figure}[!b]
    \centering
    \vspace{-9pt}
    \includegraphics[width=0.95\linewidth]%
    {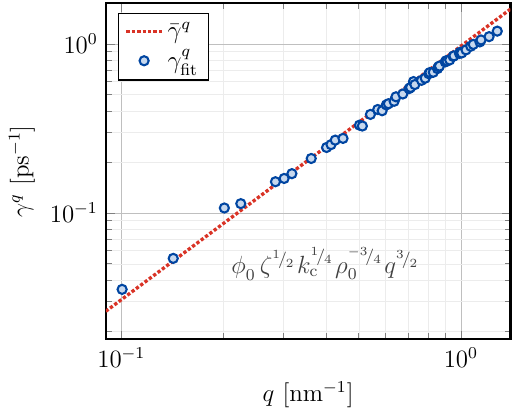}
    \vspace{-3pt}
    \caption{%
        Oscillation frequency $ \gammaqfit $ extracted from \tvacfs as a function of wavevector magnitude.
        The observed $ q^{3/2} $ scaling, combined with dimensional analysis, yields the form of $ \gammabarq $ in Eq.\ \eqref{eq_gammaq}.
        The membrane compressibility $ \kc $ does not enter the transverse dynamics of the linear theory, which reveals a coupling between modes at early times.
    }
    \label{fig_gamma_q}
\end{figure}

%
% *** Bridging molecular and hydrodynamic descriptions
%
\smallskip\noindent
\textsf{\textbf{Bridging molecular and hydrodynamic descriptions.}}---%
The \tvacfs obtained from \md simulations do not decay exponentially, but rather oscillate as they decay, as shown in Fig.\ \ref{fig_C_t}.
We find $ \Cq (t) $ is well-approximated by
$
    [
        A^q \cos (\gamma^q t)
        + B^q \sin (\gamma^q t)
    ] \mk \exp\{- \alpha^{\mkn q} t \}
$:
an exponentially damped sinusoid (see the \sm \cite{supplemental} for details).
While we cannot justify this functional form at present, we notice a robust scaling of the oscillation frequency as
$ \gammaqfit \sim q^\bfrac{3}{2} $
in Fig.\ \ref{fig_gamma_q}.
Assuming this dependence arises from membrane parameters, dimensional analysis reveals \cite{supplemental, barenblatt-similarity, eta-dim-an}
\begin{equation} \label{eq_gammaq}
    \gammaqfit
    \, \approx \ \gammabarq
    \, := \ \phi^{}_0 \, \zeta^\bfrac{1}{2} \, \kc^\bfrac{1}{4} \, \rho_0^\bfrac{-3}{4} \, q^\bfrac{3}{2}
    \, = \ \phi^{}_0 \, \sqrtl{\gammaqa \, \gammaqv \mk}
    ~.
\end{equation}
In Eq.\ \eqref{eq_gammaq}, $ \phi^{}_0 $ is a dimensionless $ \mco (1) $ constant that cannot be determined via scaling; after calculating $ \zeta $ we find
$ \phi^{}_0 \approx \bfrac{1}{3} $.
We also introduced
$
    \gammaqa
    := \cs \mk q
     = (\kc / \rhoz)^{1/2} \mk q
$
as the acoustic (sound wave) frequency, and
$ \gammaqv := \zeta \mk q^2 / \rho_0 $
as the viscous frequency of momentum diffusion.
The appearance of $ \kc $, through the sound speed $ \cs $, in oscillations of the transverse dynamics indicates longitudinal and transverse modes are coupled at early times.
In our \dopc simulations,
$ \rhoz = 3.94 \cdot 10^{-9}~\text{pg}/\text{nm}^2 $
and we find
$ \kc = 195~\text{pN}/\text{nm} $,
for which
$ \cs = 0.222~\text{nm}/\text{psec} $
\cite{supplemental}.

% begin figure
% end figure

Since the hydrodynamic prediction in Eq.\ \eqref{eq_Ch} is inconsistent with our \md results, a new procedure is required to extract the membrane viscosity.
To this end, we begin with a molecular description of transverse velocity fluctuations.
By applying the Mori--Zwanzig projection method \cite{mori-pr-1958, zwanzig-jcp-1960, zwanzig-pr-1961, mori-ptp-1965}, we find the evolution equation for $ \Cq $ is given by \cite{evans-morriss}
\begin{equation} \label{eq_Cm}
    \dd{\Cq (t)}{t}
    \, + \, \int_0^t \Kq (t - t') \, \Cq (t') ~ \td t'
    \, = \, 0
    ~.
\end{equation}
Equation \eqref{eq_Cm} is microscopically exact---though the analytical form of the memory kernel $ \Kq (t) $ is unknown.
In the hydrodynamic limit, dynamics are assumed to be Markovian: substituting
$ \Kqh := \xiq \mk \delta (t) $,
where $ \delta (t) $ is the Dirac delta function, into Eq.\ \eqref{eq_Cm} yields Eq.\ \eqref{eq_Ch}.
The microscopic and hydrodynamic perspectives are thus connected by collapsing the short-time, molecular features of $ \Kq (t) $ onto $ \Kqh (t) $---for which \cite{balucani-pra-1987, balucani}
\begin{equation} \label{eq_xiq_Kq}
    \xiq
    \, = \, \int_0^\infty \Kq (t) ~ \td t
    ~.
\end{equation}
While the analytical form of $ \Kq $ is not known, we evaluate the right-hand side of Eq.\ \eqref{eq_xiq_Kq} as follows.
First, we let
$ \tilde{f} (s) := \int_0^\infty \mkn e^{-s t} f(t) \mk \td t $
denote the Laplace transform; $ \xiq $ can then be expressed as $ \tilKq (s = 0) $.
Next, taking the Laplace transform of Eq.\ \eqref{eq_Cm} and recognizing $ \Cq (0) = 1 $ by construction yields
$ \tilCq (s) = 1 / ( s + \tilKq (s) ) $,
for which
$ \tilKq (0) = 1 / \tilCq (0) $.
Finally, $ \tilCq (0) $ is the time integral of $ \Cq (t) $, which can be determined from the \md data.
Equation \eqref{eq_xiq_Kq} is thus equivalently given by
$ \xiq = 1 / \tilCq (0) $,
for which
\begin{equation} \label{eq_zeta_q}
    \zeta (q)
	\, = \, \dfrac{\rhoz / q^2}{\tilCq (0)}
	\, - \, \dfrac{2 \mk \mu}{q} \, \tanh \big( \ellz \mk q \big)
\end{equation}
according to Eq.\ \eqref{eq_xiq}.
In Eq.\ \eqref{eq_zeta_q}, $ \zeta (q) $ is the wavevector-dependent membrane shear viscosity, within the language of generalized hydrodynamics \cite{evans-morriss, hansen}.
Once $ \zeta (q) $ is calculated from \md simulations, it is extrapolated to small $ q $: a Taylor expansion about
$ q = 0 $,
combined with symmetry arguments, reveals
$ \zeta(q) \approx \zeta (1 - b \mk q^2) $
\cite{hansen, evans-morriss};
here $ \zeta $ is the same transport coefficient entering the continuum equations.

% begin figure
\begin{figure}[!b]
    \centering
    \vspace{-9pt}
    \includegraphics[width=0.95\linewidth]{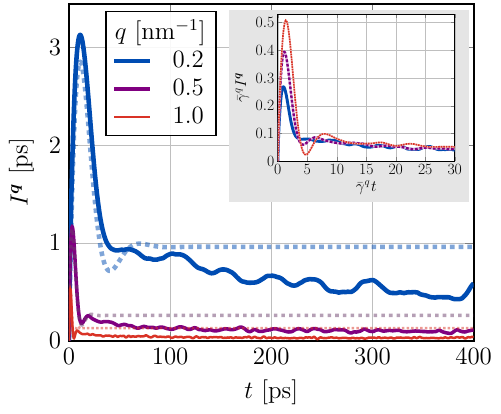}
    \vspace{-3pt}
    \caption{%
        Running integral $ \Iq (t) $ of the normalized \tvacfs shown in Fig.\ \ref{fig_C_t} (solid lines).
        The slow decrease at long times is not captured by the damped sinusoidal fits of $ \Cq (t) $, whose running integrals are shown as dashed lines.
        (inset) A common structure across wavelengths is revealed when $ \Iq $ and $ t $ are scaled with the characteristic frequency $ \gammabarq $.
    }
    \label{fig_I_t}
\end{figure}

\begin{figure*}[!b]
    \centering
    \includegraphics[width=0.45\linewidth]{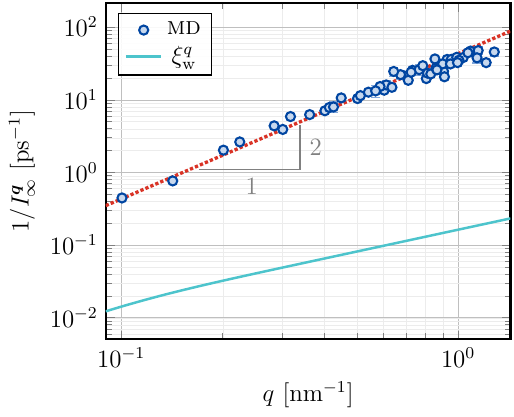}
    \hspace{20pt}
    \includegraphics[width=0.45\linewidth]{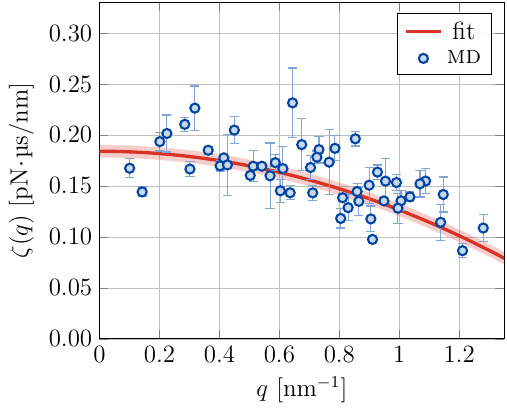}
    \begin{tikzpicture}[overlay,remember picture]
		\node at (-16.8 , 6.4) {\large(a)};
		\node at ( -7.9 , 6.4) {\large(b)};
	\end{tikzpicture}
    \vspace{-4pt}
    \caption{%
        Protocol to determine the shear viscosity of a \dopc bilayer in the fluid phase at 25$^\circ \,$C.
        (a) The Einstein--Helfand method is used to calculate $ \Iqinf $ from \md simulations.
        It is convenient to plot $ 1 / \Iqinf $, which equals the decay rate $ \xiq $.
        The contribution to $ \xiq $ from the surrounding water,
        $ \xiqw :=  (2 \mk \mu \mk q / \rhoz) \tanh (\ellz q) $,
        is negligible over the wavenumbers considered---as shown by the solid cyan line \cite{mu-water}.
        Consequently,
        $ \xiq \approx q^2 \mk \zeta(q) / \rhoz $
        and $ 1 / \Iqinf $ roughly scales as $ q^2 $ (dashed red line).
        (b) The generalized membrane shear viscosity is determined from \md data according to Eq.\ \eqref{eq_zeta_q}, as shown by the blue circles.
        Results are fit to
        $ \zeta(q) = \zeta (1 - b \mk q^2) $,
        as shown by the red line; the thick bands depict one standard error in $ \zeta $.
        In both plots, error bars in the \md data depict one standard error, as determined with the jackknife method (see the \sm \cite{supplemental} for details).
    }
    \label{fig_zeta_q}
\end{figure*}
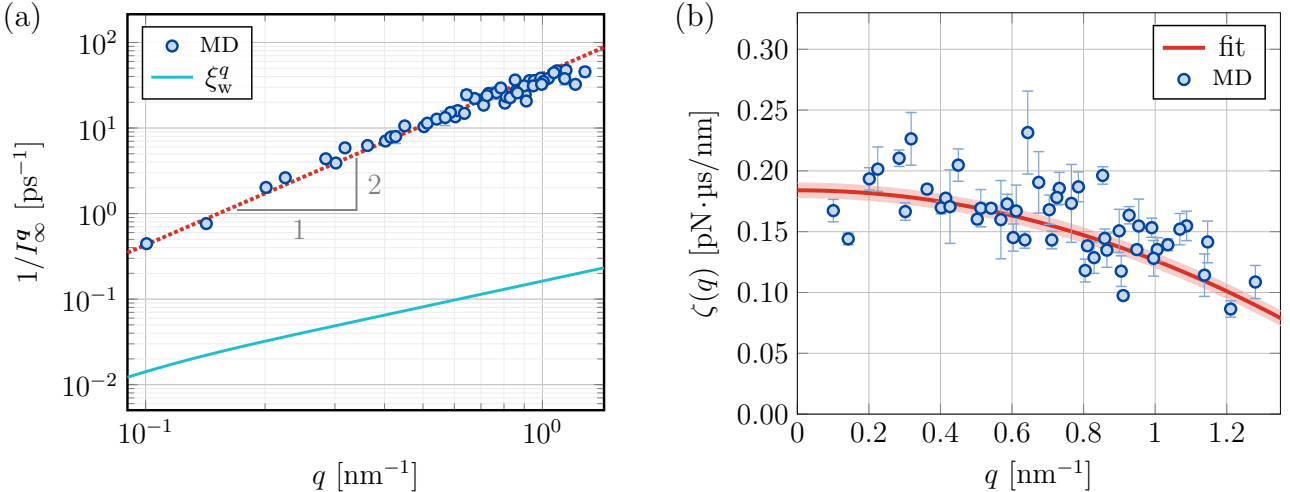

%
% *** Calculating the membrane viscosity
%
\smallskip\noindent
\textsf{\textbf{Calculating the membrane viscosity.}}---%
To calculate $ \zeta (q) $ from Eq.\ \eqref{eq_zeta_q}, we require the infinite-time integral
$ \Iqinf := \int_0^\infty \mkn \Cq (t) \, \td t $.
An \md trajectory instead provides the run-\\[-2pt]\noindent%
ning integral 
$ \Iq (t) := \int_0^{t} \Cq (t') \mk \td t' $.
Figure \ref{fig_I_t} shows that at long times, $ \Iq (t) $ continues to vary slowly and is affected by statistical noise---making its inferred asymptote sensitive to the chosen cutoff time $ \ttc $.
Nevertheless, the curves in Fig.\ \ref{fig_I_t} exhibit qualitatively similar behavior when $ t $ and $ \Iq $ are non-dimensionalized with the characteristic frequency $ \gammabarq $, suggesting a common structure in the relaxation.
This similarity does not, however, justify extrapolating $ \Iqinf $ from the fits in Fig.\ \ref{fig_C_t}: although those fits describe early-time oscillations, they do not capture the gradual decrease of $ \Iq (t) $ for
$ \gammabarq \mk \ttc \gtrsim 2 \pi $
(see Fig.\ \ref{fig_I_t}).
Hence, $ \Iqinf $ cannot be determined by either fitting $ \Cq (t) $
at early times, or approximating it as
$ \Iqinf \approx \Iq (\ttc) $
for an arbitrary cutoff time $ \ttc $.

We calculate $ \Iqinf $ with the Einstein--Helfand method \cite{helfand-pr-1960, allen-tildesley}, which estimates the infinite-time integral as the slope of a linearly growing moment (see the \sm \cite[\S 2.1]{supplemental} for details).
Briefly, we introduce
$ \hatrqperp (t) := \int_0^t \hatvqperp (t') \, \td t' $
as the time integral of the transverse velocity.
Next, the Einstein--Helfand moment
$ \Gq := \langle \mk \hatrqperp (t) \, \hatrnqperp (t) \mk \rangle $
is found to be given by
$
    \Gq (t)
    = 2 \mk t \int_0^t \hatCqv (t') \, \td t'
    - 2 \int_0^t t' \mk \hatCqv (t') \, \td t'
    \!
$,
where $ \hatCqv $ is the un-normalized \tvacf.
As
$ t \rightarrow \infty $,
the first term on the right-hand-side above dominates the second, and
$
    \Gq (t)
    \sim
    2 \mk t \, \Iq \mkn (t) \, \langle \mk \hatvqperp (0) \, \hatvnqperp (0) \mk \rangle
$.
The long-time limit of $ \Iq (t) $, which approximates $ \Iqinf $, is thus extracted from the slope of $ \Gq $ versus $ t $ at long times.
Since $ \Gq $ is fit to a straight line over a time window, the error in our estimate is improved relative to the approximations of $ \Iqinf $ discussed above.

Figure \ref{fig_zeta_q} depicts our revised protocol to calculate the shear viscosity $ \zeta $ of the \dopc bilayer characterized in Figs.\ \ref{fig_C_t}--\ref{fig_I_t}.
Long-time integrals $ \Iqinf $ are calculated with the Einstein--Helfand method, and $ 1 / \Iqinf $ is plotted as a function of $ q $ in Fig.\ \figpart{fig_zeta_q}{a}.
Recalling
$ 1 / \Iqinf = \xiq $,
we observe
$ (2 \mk \mu \mk q / \rhoz) \tanh (\ellz q) \ll \xiq $,
for which water contributes negligibly to the total drag on transverse modes [cf.\ Eq.\ \eqref{eq_xiq}].
As a consequence,
$ 1 / \Iqinf \sim q^2 $,
as shown by the dashed red line in Fig.\ \figpart{fig_zeta_q}{a}.
Figure \figpart{fig_zeta_q}{b} shows the generalized membrane shear viscosity $ \zeta (q) $ as determined from $ \Iqinf $ data via Eq.\ \eqref{eq_zeta_q}.
Following prior investigations of bulk systems \cite{palmer-pre-1994, balucani-pre-2000, hess-jcp-2002}, we fit our \md calculation of $ \zeta (q) $ to the expected form
$ \zeta \, ( 1 \mk - \mk b \mk q^2 ) $ to find
$
    \zeta^{}_{\dopc}
    = 0.184
$
$
    \pm \, 0.006
    ~ \text{pN} \! \cdot \! \text{\textmu s}/\text{nm}
$
and
$
    b^{}_{\dopc}
    = 0.31
    \pm 0.07
    ~ \text{nm}^2
$%
---shown by the red parabola in Fig.\ \figpart{fig_zeta_q}{b}.
In \S3 of the \sm \cite{supplemental}, we similarly analyze a \dppc bilayer at 67$^\circ \,$C to obtain
$
    \zeta^{}_{\dppc}
$
$
    = 0.064
    \pm \, 0.005
    ~ \text{pN} \! \cdot \! \text{\textmu s}/\text{nm}
$
and
$
    b^{}_{\dppc}
    = 0.18
    \pm 0.08
    ~ \text{nm}^2
$.
The agreement of the $ q $-dependence of $ \Iqinf $ and $ \zeta (q) $ with our theoretical predictions, over a decade of wavevectors and for two distinct bilayers, provides strong support for the \mir methodology.

%
% *** Non-equilibrium validation
%
\smallskip\noindent
\textsf{\textbf{Non-equilibrium validation.}}---%
We next verify the \mir result against an independent measurement of the membrane viscosity from nonequilibrium \md (\nemd) simulations, in which a spatially-varying external force
$ \bmFjext \mkn = \Fbar \cos(\qbar \mk y^{}_j) \mk \bmex $
is applied to each lipid molecule.
Here $ j $ is the molecule index,
$ \qbar = 2 \mk \pi / \elly $,
we choose the value of $ \Fbar $,
and $ y^{}_j $ is the $ y $-component of the molecular center-of-mass.
Upon coarse-graining, we find the additional body force per area 
$
    \bmfext (\bmx)
    \mkn = (\Fbar \rhoz / m) \cos(\qbar \mk y) \, \bmex
$
enters the continuum membrane equations, where $ m $ denotes the mass of a lipid molecule.
We then solve for the steady-state membrane velocity as
$
    \bm{\vbar} (\bmx)
    \mkn = \vbar \cos (\qbar \mk y) \, \bmex
$,
where
$ \vbar = \Fbar / (\xiqbar \mk m) $
\cite{supplemental}.
Importantly, we find that $ \zeta $---through the rate $ \xiqbar $---modulates the magnitude of the membrane velocity in response to the external force.

The proportionality constant $ \xiqbar \mk m $ between $ \vbar $ and $ \Fbar $ is now calculated in \nemd simulations, from which $ \zeta $ is determined via Eq.\ \eqref{eq_xiq}.
To this end, we choose a value of $ \Fbar $ and simulate until a steady state is reached.
Next, we project the instantaneous microscopic membrane velocity field onto the mode $ \cos (\qbar \mk y) \mk \bmex $ and average about the nonequilibrium steady state to obtain $ \vbaravg $, as described in the \sm \cite[\S4]{supplemental}.
This procedure is repeated for multiple values of $ \Fbar $, and we look for $ \vbaravg / \Fbar $ to reach a plateau at small $ \Fbar $---which yields $ \xiqbar \mk m $.
The membrane viscosities
$
    \zeta^{\textsc{nemd}}_{\dopc}
    = 0.26 \pm 0.01
$
pN$ \mk \cdot \mk $\textmu s$ \mk / $nm
and
$
    \zeta^{\textsc{nemd}}_{\dppc}
    =
$
$
    0.10 \pm 0.01
$
pN$ \mk \cdot \mk $\textmu s$ \mk / $nm
are thus obtained from Eq.\ \eqref{eq_xiq}; both overshoot their \mir counterparts by 40--60\%.
This systematic discrepancy is attributed to the narrowness of the linear-response regime, and the difficulty of thermostatting a system into which momentum is continually injected \cite{evans-morriss}. 
The \nemd simulations thus serve only as an order-of-magnitude benchmark, and in this respect are consistent with our \mir results.

%
% *** Discussion and conclusions
%
\smallskip\noindent
\textsf{\textbf{Discussion and conclusions.}}---%
In this Letter, we developed the general method of molecular interfacial rheology (\mir), with which surface viscosities of molecular interfaces are extracted from equilibrium \md simulations.
Hydrodynamic couplings between the 2D surface and surrounding medium were captured by our continuum description, and connected to the decay of molecular velocities through the Mori--Zwanzig formalism.
Challenges associated with \pbcs in the $ z $-direction were overcome by positioning the system between solid walls.
The shear viscosity could then be calculated from the time integral of transverse velocity correlations, without an assumed form of the microscopic relaxation.
For fluid-phase \dopc and \dppc bilayers, the predicted wavevector dependence of $ \zeta (q) $ was observed over a decade in $ q $.
Independent \nemd calculations also yielded consistent viscosities, and validated our \mir developments.
%\as{These successes motivate the use of solid walls in future molecular investigations.}

Beyond the specific values of $ \zeta $ that were determined, \mir is useful in several ways.
First, with experimental measurements of the membrane viscosity spanning several orders of magnitude \cite{honerkamp-prl-2013, nagao-prl-2021, faizi-bpj-2022, suja-prf-2025}, our findings ($i$) are a quantitative benchmark for such measurements and ($i \mkn i$) allow for the systematic investigation of how $ \zeta $ depends on e.g.\ temperature and composition.
In addition, \mir offers \md developers a dynamic observable against which membrane force fields can be parametrized and tested.
More broadly, the methods developed here can characterize the rheology and transport of other molecular interfaces---especially in cases where surface and bulk dynamics are coupled.

%\noindent%
%For DOPC:
%
%$ \ellx = 62.48~\text{nm} $,
%
%$ \ellz = 13.41~\text{nm} $,
%
%$ \kc = 194~\text{pN}/\text{nm} $,
%
%$ \zeta = 661~\text{pN}\cdot\text{psec}/\text{nm} $,
%
%$ \rhoz = 3.938 \cdot 10^{-9}~\text{pg}/\text{nm}^2 $,
%
%$ c = 0.222~\text{nm}/\text{psec} $.

% *** ACKNOWLEDGMENTS *** %
\small
\smallskip\balance

%
% *** Acknowledgments
%
\medskip
\noindent\textbf{\textsf{Acknowledgments.}}
It is a pleasure to thank
Prof.\ \href{https://mandadapu-group.github.io}{Kranthi Mandadapu}
for many discussions regarding membrane viscosities, starting with Refs.\ \cite{sahu-pre-2017, sahu-pre-2020}.
We are grateful to Prof.\ \href{https://sites.utexas.edu/ganesan/}{Venkat Ganesan}
for helpful questions and comments about our calculations.
We thank Prof.\ \href{https://chemistry.berkeley.edu/people/david-limmer}{David Limmer} for insightful discussions, recommending the \nemd validation, and bringing several references to our attention.

This work was partially supported by the \href{https://welch1.org/}{Welch
Foundation} via Grant No.\ F-2208.
We are grateful to the \href{https://tacc.utexas.edu}{Texas Advanced Computing Cluster}, where the majority of \md simulations were carried out.

%
% *** Declaration of interests
%
%\medskip\small
%\noindent\textbf{\textsf{Declaration of interests.}}
%%
%There are no conflicts to declare.

% *** REFERENCES *** %
\small%
\subsection*{References}
\bibliographystyle{bibStyle}
\bibliography{refs}

\begin{thebibliography}{10}
\bibitem{parallel}
\bibinfo{title}{{The use of `$\parallel$' as a subscript here is unrelated to
  its use in the operator $ \bmnablapara $.}}
\bibitem{eta-dim-an}
\bibinfo{title}{{The dilational viscosity $ \eta $ could also be included.
  Since $ \eta $ and $ \zeta $ have the same dimensions, an arbitrary function
  of $ \eta / \zeta $ would then enter the relation for $ \gammaq $. However,
  the appearance of $ \eta $ in the transverse dynamics again indicates
  transverse--longitudinal mode coupling at early times.}}
\end{thebibliography}

\end{document}